\documentclass[12pt,preprint]{aastex}

\usepackage{graphicx}
\usepackage{amssymb}
\slugcomment{ {\bf  Revised, 05 Apr. 2005} }

\begin{document}

\title{ The Mauna Kea Observatories Near-Infrared Filter Set.   III. Isophotal Wavelengths and Absolute Calibration  }

\author{A. T. Tokunaga}
\affil{Institute for Astronomy, University of Hawaii, 2680 Woodlawn Drive, 
Honolulu, HI  96822}
\email{tokunaga@ifa.hawaii.edu}

\and

\author{W. D. Vacca}
\affil{Stratospheric Observatory for Infrared Astronomy/Universities Space Research Association, NASA Ames Research Center, MS 144-2, Moffett Field, CA 94035} \email{wvacca@mail.arc.nasa.gov}

\begin{abstract}

The isophotal wavelengths, flux densities, and AB magnitudes for Vega
($\alpha$~Lyr) are presented for the Mauna Kea Observatories
near-infrared filter set. We show that the near-infrared absolute
calibration for Vega determined by Cohen et al. and M\'{e}gessier
are consistent within the uncertainties, so that either absolute
calibration may be used.  

\end{abstract}


\section{INTRODUCTION}

Simons \& Tokunaga (2002) and Tokunaga, Simons, \& Vacca (2002)
defined a 1--5~$\mu$m filter set that is designed to maximize
sensitivity while minimizing the effects of atmospheric absorption
on reducing the signal-to-noise ratio.  This filter set is intended to provide
good transformation between observatories located at altitudes of 2--4~km. 
Filter production runs have been organized to produce these filters, which 
are in use at more than 30 institutions. This filter set provides greater
transmission than those advocated by Young, Milone, \& Stagg (1994), 
and was designed to provide nearly ideal photometric accuracy.  
To distinguish this filter set from others, we refer to it as the 
Mauna Kea Observatories near-infrared (MKO-NIR) filter set.

In this paper, we present the isophotal wavelengths, flux densities,
and AB magnitudes for Vega for the MKO-NIR filter set.  We 
compare the absolute calibration advocated by  Cohen et al. (1992)
to that of M\'{e}gessier (1995), and we show that there is no
significant difference in the flux densities for Vega derived by these
authors.

\section{ISOPHOTAL WAVELENGTHS AND ZERO MAGNITUDE FLUX DENSITIES}

\subsection{The Definition of Isophotal Wavelength}

The number of  photo-electrons detected per second from a
source with an intrinsic spectral energy distribution
$F_\lambda(\lambda)$ is given by
\begin{eqnarray}
\label{eq:pdef}
N_p & = & \int F_\lambda(\lambda) S(\lambda)/ h\nu \ d\lambda \\
\label{eq:fdef}
         & = & \frac{1}{hc}\int \lambda F_\lambda(\lambda) S(\lambda) d\lambda
\end{eqnarray}
where  $S(\lambda)$ is  the total system response given by 
\begin{equation}
S(\lambda) = T(\lambda) Q(\lambda) R(\lambda)  A_{\rm{tel}}  ~~.
\end{equation}  
Here $T(\lambda)$ is the atmospheric transmission, $Q(\lambda)$ is the
product of the throughput of the telescope, instrument, and quantum
efficiency of the detector, $R(\lambda)$ is the filter response
function, and $A_{\rm{tel}}$ is the telescope collecting area.  The system 
response $S(\lambda)$ is equal to the relative spectral response (RSR) 
defined by Cohen et al. (2003).

If $F_\lambda(\lambda)$ and $S(\lambda)$ are both continuous and
$S(\lambda)$ is nonnegative over the wavelength interval, then from
equation (2) and the mean value theorem for integration there exists
a $\lambda_{\rm{iso}}$ such that 
\begin{equation}
F_\lambda(\lambda_{\rm{iso}})   \int \lambda S(\lambda) d\lambda  = 
\int \lambda  F_\lambda(\lambda)  S(\lambda) d\lambda  ~~.
\end{equation}
Rearranging this, we obtain
\begin{equation}
F_\lambda(\lambda_{\rm{iso}})  =  \langle F_\lambda \rangle  =  
\frac{\int \lambda F_\lambda(\lambda) S(\lambda) d\lambda} {\int \lambda S(\lambda) d\lambda} ~~,
\end{equation}
where $\lambda_{\rm{iso}}$ denotes the ``isophotal wavelength'' and $\langle
F_\lambda \rangle$ denotes the mean value of the intrinsic flux above the
atmosphere (in units  of W m$^{-2}$ $\mu$m$^{-1}$) over the wavelength interval
of the filter.
Thus $\lambda_{\rm{iso}}$ is the wavelength at which the monochromatic flux
$F_\lambda(\lambda_{\rm{iso}}) $ equals the mean flux in the passband. Hence
$\lambda_{\rm{iso}}$ and $F_\lambda(\lambda_{\rm{iso}}) $ are the wavelength and
monochromatic flux density, respectively, that best represent a broadband
(heterochromatic) measurement.  In addition we choose to use isophotal wavelengths 
for consistency with the extensive series of papers on infrared calibration by Cohen
and collaborators.

In a similar fashion,
\begin{equation}
F_\nu(\nu_{\rm{iso}})  =  \langle F_\nu \rangle  =  
\frac{ \int  F_\nu(\nu) S(\nu)/\nu \ d\nu} {\int S(\nu)/\nu \ d\nu} ~~,
\end{equation}
where $\nu_{\rm{iso}}$ denotes the ``isophotal frequency'' and 
$\langle F_\nu \rangle$ denotes the mean value of the intrinsic flux 
above the atmosphere (in units W m$^{-2}$ Hz$^{-1}$) 
over the frequency interval of the filter.

We show in Table 1 the MKO-NIR filter isophotal wavelengths for Vega.
These were calculated from equation (5) using an atmospheric model of
Vega computed by R. Kurucz\footnote{http://kurucz.harvard.edu/stars/VEGA} 
with the parameters $T$$_{\rm eff}$ = 9550 K, log($g$) = 3.95, 
$v_{\rm{turb}}$ = 2 km s$^{-1}$,  $v_{\rm{rot}}$ = 25 km s$^{-1}$, 
and [Fe/H] = $-$0.5. These parameters are the same 
as those adopted by Bohlin \& Gilliland (2004).  The model has a resolving power 
of $10^5$ and has been scaled to the absolute flux level of  
$3.46 \times 10^{-9}$ erg cm$^{-2}$ s$^{-1}$ \AA$^{-1}$  at 5556 \AA\  as 
determined by M\'{e}gessier (1995).  We used ATRAN (Lord 1992) to calculate
the atmospheric transmission $T(\lambda)$ for a range of precipitable water
values between 0 and 4 mm, an airmass of 1.0, and the altitude of Mauna
Kea. The measured filter response curves\footnote{
http://irtfweb.ifa.hawaii.edu/Facility/nsfcam/hist/newfilters.html}
given by Tokunaga et al.\ (2002) were used for $R(\lambda)$ (see their Fig. 1). 
We assumed that the throughput term $Q(\lambda$) was a constant over the
wavelength integrals.   Thus our calculations are precisely correct only for 
constant detector responsivity and instrumental throughput.  However
Stephens \& Leggett (2004) find that the variations in the detector 
responsivity and instrumental throughput leads to photometric variations
of $\leq$0.01 mag.  This corresponds to variations
of $\lambda_{\rm{iso}}$ of  $\leq$0.3\%.

In accordance with the design goals of the MKO-NIR filters, variations in
$\lambda_{\rm{iso}}$ as a function of precipitable water vapor were found to be very
small ($\leq$1\%) over the range of water vapor values typically
encountered on Mauna Kea.  For the range of 0 to 4 mm of precipitable 
water vapor,  we found variations $\Delta\lambda_{\rm{iso}}$ of 0.000 $\mu$m ($J$), 
0.001 $\mu$m ($H$), 0.012 $\mu$m ($K'$), 0.013 $\mu$m ($K_s$), 0.001 
$\mu$m  ($K$), -0.014 $\mu$m ($L'$), and 0.001 $\mu$m ($M'$).

The methods employed here, in particular using the number of photons
detected and the formulation in equation (5), which includes the
wavelength terms in the integrals, are the same as those used by Cohen et
al. (1992) and subsequent papers by Cohen and his collaborators. 
For clarity we have explicitly presented the equations for calculating the isophotal
wavelengths for photon counting detectors. See Bessell et
al. (1998) for discussion of how photometric results differ between
energy measuring detectors and photon-counting detectors.

Real spectra do not necessarily satisfy the requirements of the mean
value theorem for integration, as they exhibit discontinuities. Although the 
mean value of the intrinsic flux is well-defined,  the determination of the
isophotal wavelength becomes problematic because real spectra contain
absorption lines and hence the definition can yield multiple solutions.
In addition, $\lambda_{\rm{iso}}$  is not an easily measured observational
quantity because it depends on knowledge of the intrinsic source flux
distribution $F_\lambda(\lambda)$,  which is exactly what one is
attempting to determine with broadband photometry.  Nevertheless for the 
reasons stated above we use isophotal wavelengths in this paper.
Other definitions of the filter wavelength are briefly discussed in the Appendix 
for completeness.

Any definition of the filter
wavelength suffers from the limitation that the spectral energy
distribution of the object being observed is likely to be different from
that of Vega. Since the isophotal wavelength is different for objects that
have different spectral energy distributions, a
correction factor is required to obtain the monochromatic magnitude at
the same isophotal wavelength as Vega. Hanner et al. (1984) discuss 
this problem in detail for observations of comets.


\subsection{Absolute Flux Densities for Vega}

The question of the near-infrared absolute flux densities for Vega above
the atmosphere has been discussed by Cohen et al. (1992) and
M\'{e}gessier (1995). Cohen et al. determined their absolute calibration
from a model atmosphere for Vega multiplied by the atmospheric
transmission and instrument response (filters, throughput, and filter
response).  They did not use absolute calibration measurements
by Blackwell et al. (1983) and Selby et al. (1983) because Blackwell et
al. (1990) concluded that atmospheric models of Vega offered higher
precision than the observationally determined absolute calibration in
the near-infrared. Bessell et al. (1998) also concluded that model
atmospheres are more reliable than the near-infrared absolute
calibration measurements. However M\'{e}gessier (1995) argued that the
models were not reliable and that the near-infrared absolute calibration of
Vega should be based on measurements that are independent of atmospheric
models.  Based on four model-independent measurements, M\'{e}gessier
(1995) determined an averaged absolute flux density for Vega.

We directly compared the absolute calibration of Vega determined by
Cohen et al. (1992) and M\'{e}gessier (1995). We first used a
third-order polynomial to fit the logarithm of the flux density for Vega as a
function of  the logarithm of the wavelength  from M\'{e}gessier. The
use of logarithms gives a nearly linear relationship. The wavelengths
and flux densities for 0.0 mag were taken from Table 4 in
M\'{e}gessier.  We assumed a near-infrared magnitude of 0.02 mag for 
Vega to be consistent with the visible magnitude assumed by
M\'{e}gessier.   We then  interpolated to the wavelengths cited by Cohen
et al. in their Table 1 to make direct comparisons to the M\'{e}gessier results.

The difference in the logarithm between the M\'{e}gessier and Cohen et
al. absolute flux densities for Vega are shown in Figure 1. The mean of
the difference  is $-$0.0022 $\pm$ 0.0031. Thus the absolute
flux densities of Vega as given by M\'{e}gessier and Cohen et al.  are
indistinguishable.  The uncertainty of the M\'{e}gessier and Cohen et al. flux 
densities for Vega are about 2\% and 1.45\%, respectively.

While the results discussed above show the consistency of the
independent methods of Cohen et al. and M\'{e}gessier, we note the
following caveats:

\begin{enumerate}

\item Both Cohen et al. (1992) and M\'{e}gessier (1995) rely on the
absolute calibration of Vega at 0.5556 $\mu$m.  Cohen et al. adopt a
flux density of 3.44 $\times$ 10$^{-8}$ W m$^{-2}$ $\mu$m$^{-1}$, while
M\'{e}gessier adopts 3.46 $\times$ 10$^{-8}$ W m$^{-2}$ $\mu$m$^{-1}$. 
Thus M\'{e}gessier's flux density for Vega is 0.6\% higher at $V$.

\item Recent  work by Gulliver et al. (1994) and Peterson et al. (2004) 
indicate that Vega is pole-on and a fast rotator.  Thus standard model 
atmospheres are not appropriate for Vega as discussed by Bohlin \& Gilliland 
(2004).  Nonetheless we have already shown the agreement between the 
Cohen et al. values and the model-independent results of M\'{e}gessier.
In addition, Price et al. (2004) show that the {\em Midcourse Space Experiment 
(MSX)} absolute calibration experiment is in agreement with the 
Cohen et al. (1992) values for Vega, to within the experimental errors of 1\%.

\end{enumerate}


We show in Table 1 the flux densities for Vega for the isophotal wavelengths of
the MKO-NIR filters.  We have adopted the 1--5 $\mu$m flux densities
for Vega as presented by Cohen et al. (1992) in their Table 1.  
We first computed the flux densities for Vega assuming a precipitable water 
vapor value of 2 mm at an airmass of 1.0, in addition to the parameters 
for Vega discussed in Section 2.1.    Since the model
flux density for Vega is 1.9\% lower than that used by Cohen et al. (1992), we 
increased our calculated values by 1.9\%, and this is shown in
Table 1.  Bohlin \& Gilliland (2004) also found that the flux density of Vega in the
infrared was about 2\% lower than that presented by Cohen et al. (1992).

For the $V$ isophotal wavelength and flux density calculations, we used 
the absolute spectrophotometry for Vega given by Bohlin \& Gilliland (2004).
The $V$ filter profile used was that of Landolt (1992), 
obtained from Cohen et al. (2003; see the electronic version of the paper).  
The values we obtained for this $V$ filter are shown in Table 1.

It is evident from the above discussion that there is no consistent published 
atmospheric model for Vega at both visible and infrared wavelengths.  This 
is primarily because Vega is nearly pole-on and there is 
a range of temperature from the hotter pole regions to the cooler equatorial 
regions.  A single temperature model for Vega is therefore not realistic.  
The values of isophotal wavelengths and flux 
densities shown in Table 1 represent a best estimate based on 
absolute calibrations using blackbody sources, observations of
standard stars, and atmospheric models.

We note that M\'{e}gessier (1995) found that the observed fluxes of Vega
were about 2\% higher than the atmospheric models.  This problem
was attributed to a possible near-infrared excess of Vega.  
However Leggett et al. (1986) found no infrared excess from Vega compared 
to other A0 stars.  This suggests that atmospheric models for Vega at 
near-infrared wavelengths are in error, possibly because Vega is observed
pole-on.

\subsection{Comment on the Definition of Zero Magnitude}

Infrared photometric systems at 1--5 $\mu$m are usually defined as being
based on the Johnson system or in a system in which the magnitude of
Vega is taken to be 0.0. Examples of the former include systems at the Univ. of
Arizona (Campins et al. 1985), ESO (Wamsteker 1981), SAAO (Carter 1990), 
and AAO (Allen \& Cragg 1983). In these systems, the magnitude of
Vega is defined as 0.02 or 0.03 mag. Examples of the latter include the 
systems 
CIT (Elias et al. 1982) and the Las Campanas Observatory (Persson et al.
1998).   The UKIRT photometric system (Hawarden et al. 2001; Leggett et
al. 2003) is based on the Elias et al. (1982) standard stars, so it
follows the convention that the magnitude of Vega is 0.0 mag. 
Cohen et al. (1992) adopted a magnitude of 0.0 mag at infrared 
wavelengths,  so that the flux density of Vega defines the flux density for 
0.0 mag.  This is continued in subsequent papers, and in Cohen et al. (2003) the
nonzero magnitude of Vega at optical wavelengths is taken into account.
Thus when applying the results in Table 1, one must take into consideration 
which photometric system is used.

\subsection{AB Magnitudes}

The monochromatic AB magnitudes were defined by Oke \& Gunn (1983) as 
\begin{equation}
\rm{AB} = -2.5 \log(f_\nu) - 48.60
\end{equation}
where $f{_\nu}$ is in units of ergs cm$^{-2}$ s$^{-1}$ Hz$^{-1}$ 
(see also Fukugita et al. 1996).  The constant is set so that  AB is 
equal to the $V$ magnitude for a source with a flat spectral 
energy distribution. We adopt the Vega flux densities recommended by 
Bohlin \& Gilliland (2004) (their alpha\_lyr\_stis\_002.fits file).  The visible 
flux values are tied to a flux density of 3.46 $\times$ 10$^{-8}$ W m$^{-2}$
$\mu$m$^{-1}$ at 0.5556 $\mu$m following M\'{e}gessier (1995). 
The isophotal wavelength at $V$ is 5546 \AA\footnote{Should be 
5450 \AA, see erratum.},   and the isophotal flux density
is 3.63 $\times$ 10$^{-20}$ ergs s$^{-1}$ cm$^{-2}$ Hz$^{-1}$ 
(Table 1).  A $V$ magnitude of 0.026 from Bohlin \& Gilliland (2004) is 
assumed.  Then
\begin{equation}
\rm{AB} = -2.5 \log(f_\nu) - 48.574
\end{equation}
or for F$_\nu$ expressed in units of Jy, we have
\begin{equation}
\rm{AB} = -2.5 \log(F_\nu) + 8.926  ~~~.
\end{equation}

The AB magnitudes for Vega were calculated from equation (9) and are 
shown in the AB magnitude column of Table 1. A great advantage of AB 
magnitudes is that the conversion to physical units at
all wavelengths can be obtained with a single equation:
\begin{equation}
F_{\nu} = 3720 \, 10^{-0.4 AB}  ~~~.
\end{equation}
The constant in equation 8 differs from that of Fukugita et al. (1996) because 
for Vega we assumed a different flux density value at $V$ and 
adopted a different visual magnitude.   However, it is within the uncertainty
of the absolute calibration of 2\%  for Vega stated by Oke \& Gunn (1983).

\section{SUMMARY}

\begin{enumerate}

\item Isophotal wavelengths, flux densities, and AB magnitudes for
Vega are derived for MKO-NIR filter set. 

\item The absolute calibrations by Cohen et al. (1992) and 
M\'{e}gessier (1995) are shown to be identical within the uncertainties.  
We adopt the 1--5 $\mu$m absolute calibration of Cohen et al. to be 
consistent with the subsequent papers by Cohen and his colleagues.

\item The $V$-band isophotal wavelength and flux density is given for 
completeness using the renormalized Vega model and the absolute calibration
adopted by Bohlin \& Gilliland (2004).  The constant in the AB magnitude
definition was determined from the recent {\em Hubble Space Telescope (HST)} measurements of the $V$ magnitude and  flux density of Vega and
differs from that defined by Oke \& Gunn (1983).

\item There is no self-consistent atmospheric model for Vega at visible
and infrared wavelengths.  Improvements to Table 1 can be expected with 
models that take into account that Vega is observed pole-on, and also with  observations of a grid of A0 stars (including Sirius) to eliminate the dependence 
on currently unreliable atmospheric models for Vega in the infrared.

\end{enumerate}

\acknowledgments{We thank M.\ Cohen, D. Peterson, and T. Nagata for useful 
discussions and Steve Lord for making ATRAN available to us. A.T.T. was supported 
by NASA
Cooperative Agreement number NCC 5-538. This research has made use of
NASA's Astrophysics Data System Bibliographic Services.  
}

\section{APPENDIX A}

Other definitions for effective wavelengths are briefly discussed here. 
A more detailed discussion may be found in Golay (1974). In a manner similar 
to that used in equation (4), we can define the ``effective wavelength'' as 
%
%
$$\lambda_{\rm{eff}}  \int F(\lambda) S(\lambda) d\lambda  = 
\int \lambda  F_\lambda(\lambda) S(\lambda) d\lambda  ~~,    \eqno(\rm{A1}) $$
%
and hence
%
$$\lambda_{\rm{eff}}  = \frac{\int \lambda F_\lambda(\lambda) S(\lambda) d\lambda}{\int F_\lambda(\lambda) S(\lambda) d\lambda}   ~~~.   \eqno(\rm{A2}) $$
%

This is the wavelength at which the flux distribution in energy units,
integrated over the passband and then converted to photons, equals the
flux distribution in photon units integrated over the passband.
Alternatively, $F_\lambda(\lambda) S(\lambda)$ can be thought of as a
probability distribution for the detection of energy from the source,
and therefore, $\lambda_{\rm eff}$ is the mean wavelength of the
passband as weighted by the energy distribution of the source over the
band. In a similar manner, we can determine a mean wavelength for the
passband as weighted by the photon distribution of the source over
the band. In this case, 

$$\lambda'_{\rm eff} = \frac{\int
\lambda^2 F_\lambda(\lambda) S(\lambda) d\lambda}{\int \lambda
F_\lambda(\lambda) S(\lambda) d\lambda}   ~~~.    \eqno(\rm{A3}) $$

Here, $\lambda F_\lambda(\lambda) S(\lambda)$ can be thought of as the
probability distribution for detecting a photon from the source. Note
that both effective wavelength definitions depend on the spectral energy
distribution of the source.

We can define a source-independent wavelength as follows:
%
$$\int P_\lambda(\lambda) S(\lambda) d\lambda  = 
\int  F_\lambda(\lambda) S(\lambda) / h \nu \, d\lambda ~~,     \eqno(\rm{A4}) $$
%
where $P_\lambda(\lambda) = F_\lambda(\lambda) / h\nu$ is the photon
flux from the object.   Thus, 
$$ \langle P_\lambda \rangle \int S(\lambda) d\lambda  =   
 \frac{\langle F_\lambda \rangle}{hc} \int \lambda S(\lambda) d\lambda  ~~~.
    \eqno(\rm{A5}) $$
%
Setting  
$ \langle P_\lambda \rangle = 
\lambda_{\rm{0}} \langle F_\lambda \rangle / hc$ yields 
%
$$\lambda_0 = \frac{\int \lambda S(\lambda) d\lambda}{\int S(\lambda) d\lambda} ~~,
   \eqno(\rm{A6}) $$
%
which is  the ``mean wavelength'' of the system. 

We can express equation (1) in terms of frequency to derive another filter
wavelength.  From equation (1) we have
%
$$N_p =  \frac{1}{hc}\int \lambda F_\lambda(\lambda) S(\lambda) d\lambda  =   \frac{1}{h}\int F_\nu(\nu) S(\nu) d\nu / \nu  ~~~.    \eqno(\rm{A7}) $$
%
In a manner similar to the derivation given above for equation (14),
%
$$\frac{1}{c} \langle F_\lambda \rangle \int \lambda S(\lambda) d\lambda 
 =  \langle F_\nu \rangle \int \frac{S(\nu)}{\nu} d\nu     \eqno(\rm{A8}) $$
%
$$ =  \langle F_\nu \rangle  \int \frac{S(\lambda)}{\lambda} d\lambda  ~~,
   \eqno(\rm{A9}) $$
%
where the last equation results from the fact that $d\nu / \nu  =  d\lambda / \lambda$. 
If we set 
%
$$\langle F_\nu \rangle  = 
\langle F_\lambda \rangle \lambda_{\rm{pivot}}^2/c  ~~,     \eqno(\rm{A10}) $$
%
we obtain
%
$$\lambda_{\rm{pivot}} = \sqrt{\frac{\int \lambda S(\lambda) d\lambda}{\int S(\lambda) d\lambda/ \lambda}}  ~~,     \eqno(\rm{A11}) $$
%
which is known as the ``pivot wavelength'' of the system. The pivot
wavelength provides an exact relation between $F_{\nu}$ and
$F_{\lambda}$ given by equation (A10). This definition is used in the
$HST$ Synphot Users Guide (Bushouse \&  Simon 1998; see also 
Koorneef et al. 1986).

The wavelengths based on the different definitions are shown in Table 2.


\begin{figure}
\epsscale{1.0}
plotone{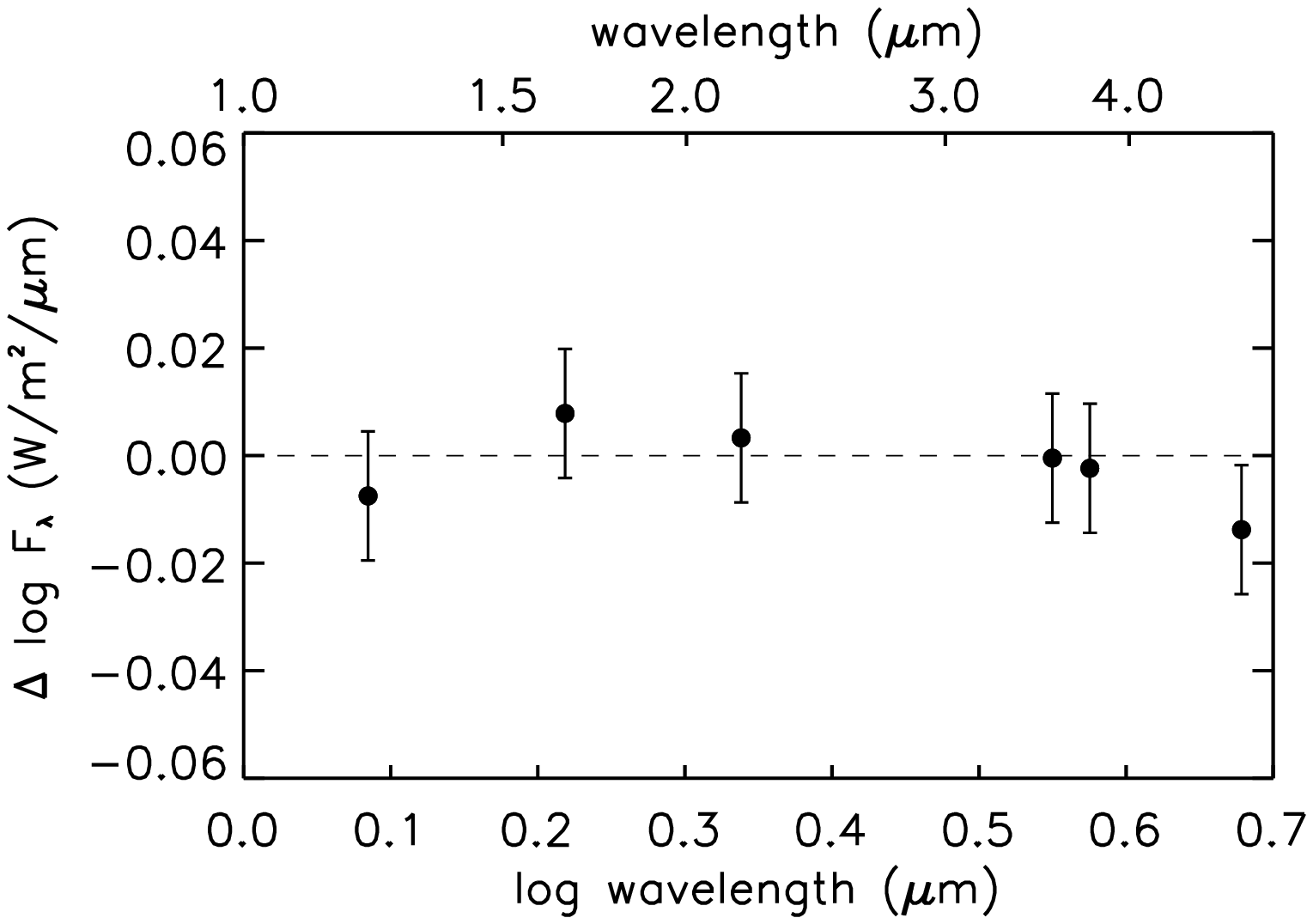}
\vspace{-2.0in}
\hspace{-1.25in}
\caption{ Difference of the logarithm of the Cohen et al. (1992) and
M\'{e}gessier (1995)  flux densities for Vega.  The M\'{e}gessier flux
density was fitted with a third order polynomial and 
subtracted from the Cohen et al.  values (see text).
}
\end{figure}

\clearpage

\begin{deluxetable}{ccccc}
\tablecaption{Isophotal wavelength, flux densities, and AB magnitudes for Vega}
\tablewidth{0pt}
\tablehead{
\colhead{Filter} & \colhead{$\lambda_{\rm{iso}}$} & \colhead{F$_\lambda$} & \colhead{F$_\nu$}  & \colhead{AB magnitude } \\
\colhead{ }   & \colhead{($\mu$m)} & \colhead{(W m$^{-2}$ $\mu$m$^{-1}$)} & \colhead{(Jy)} &  \colhead{ } \\
}
\startdata

V	 &0.5446	&3.68E-08	&3630	&0.026 \\
J	&1.250	&3.01E-09	&1560	&0.943 \\
H	&1.644	&1.18E-09	&1040	&1.38 \\
K$'$	&2.121	&4.57E-10	&686	&1.84 \\
Ks	&2.149	&4.35E-10	&670	&1.86 \\
K	&2.198	&4.00E-10	&645	&1.90 \\
L$'$	&3.754	&5.31E-11	&249	&2.94 \\
M$'$	&4.702	&2.22E-11	&163	&3.40 \\

 \enddata

\tablecomments{
We assume the Landolt V filter profile (see text) and 2mm of precipitable water.}

\end{deluxetable}

\begin{deluxetable}{cccccc}

\tablecaption{Isophotal, Effective, Mean, and Pivot Wavelengths for the 
MKO-NIR filters}
\tablewidth{0pt}
\tablehead{
\colhead{Filter}  & \colhead{ $\lambda_{\rm{iso}}$}   &   
\colhead{$\lambda_{\rm{eff}}$}  &\colhead{ $\lambda'_{\rm{eff}}$} &  
\colhead{$\lambda_{\rm{0}}$}  &  \colhead{$\lambda_{\rm{pivot}}$}  \\
\colhead{ }   & \colhead{($\mu$m)} &  \colhead{($\mu$m)}  & 
\colhead{($\mu$m)}  & \colhead{($\mu$m)}  & \colhead{($\mu$m)} \\
}

\startdata
$J$    &   1.250  &  1.241  & 1.243  & 1.248  &  1.247  \\
$H$    &   1.644  &  1.615  & 1.619  & 1.630  &  1.628  \\
$K'$   &   2.121  &  2.106  & 2.111  & 2.123  &  2.121  \\
$Ks$   &   2.149  &  2.138  & 2.141  & 2.151  &  2.150  \\
$K$    &   2.198  &  2.186  & 2.190  & 2.202  &  2.200  \\
$L'$   &   3.754  &  3.717  & 3.727  & 3.757  &  3.752  \\
$M'$   &   4.702  &  4.680  & 4.681  & 4.684  &  4.684  \\
\enddata

\end{deluxetable}

\section{ERRATUM -  to be submitted}

The isophotal wavelength for the V filter given in Section 2.4 and Table 1 should be
5450 \AA.

We did not show the isophotal frequency in Table 1 that should be
used with the flux density in frequency units (Jy) and the AB magnitudes derived
from it.  Therefore we present in Table 3 the isophotal frequencies.  Note that
the isophotal wavelength and isophotal frequency must be derived using equations 
(5) and (6) since $\lambda_{iso} \neq c / \nu_{iso}$.  In addition, F$_\nu$ 
and AB magnitudes should always be plotted with $\nu_{iso}$.

\begin{deluxetable}{cccc}
\tablecaption{Isophotal frequency, flux densities, and AB magnitudes for Vega}
\tablewidth{0pt}
\tablehead{
\colhead{filter} & \colhead{$\nu_{\rm{iso}}$} & \colhead{F$_\nu$} &  \colhead{AB mag} \\
\colhead{name}   & \colhead{($\times$10$^{14}$ Hz)} & \colhead{(Jy)} &  \colhead{(mag)} \\
}
\startdata

V      &  5.490  &  3630   &  0.026  \\
J       &  2.394  &  1560   &  0.941  \\
H      &  1.802  &  1040   &  1.38    \\
K$'$  &  1.413  &  686    &   1.84   \\
Ks     &  1.395  &  670    &  1.86    \\
K       &  1.364  &  645    &  1.90    \\
L$'$  &  0.7982  &  249    &  2.93    \\
M$'$	 &  0.6350  &  163    &  3.40    \\

 \enddata


\end{deluxetable}


\begin{thebibliography}{}

\bibitem[]{} Allen, D. A., \& Cragg, T. A. 1983, \mnras, 203, 777

\bibitem[]{} Bessell, M. S., Castelli, F., \& Plez, B. 1998, \aap, 333, 231

\bibitem[]{} Blackwell, D. E., Leggett, S. K., Petford, A. D., Mountain, C. M., \& Selby, M. J. 1983, \mnras, 205, 897

\bibitem[]{} Blackwell, D. E., Petford, A. D., Arribas, S., Haddock, D. J., \& Selby, M. J. 1990, \aap, 232, 396

\bibitem[]{} Bohlin, R. C., \& Gilliland, R. L. 2004, \aj, 127, 3508

\bibitem[]{} Bushouse, H., \& Simon, B.  1998, Synphot User's Guide (Baltimore: Space 
Telescope Science Institute).

\bibitem[]{} Campins, H., Rieke, G. H., \& Lebofsky, M. J. 1985, \aj, 90, 896

\bibitem[]{} Carter, B. S. 1990, \mnras, 242, 1

\bibitem[]{} Cohen, M., Walker, R. G., Barlow, M. J., \& Deacon, J. R. 1992, \aj, 104, 1650

\bibitem[]{} Cohen, M., Megeath, S. T., Hammersley, P. L., Mart\'{i}n-Luis, F., \& Stauffer, 
J. 2003, \aj, 125, 2645

\bibitem[]{} Elias, J. H., Frogel, J. A., Matthews, K., \& Neugebauer, G. 1982, \aj, 87, 1029; erratum 1982, \aj, 87, 1893

\bibitem[]{} Fukugita, M., Ichikawa, T., Gunn, J. E., Doi, M., Shimasaku, K., \& Schneider, D. P. 1996, \aj, 111, 1748

\bibitem[]{} Golay, M. 1974, Introduction to Astronomical Photometry, Vol. 41 in the Astrophysics and Space Science Library (Dordrecht: Reidel), pp. 39--46

\bibitem[]{} Gulliver, A. F., Hill, G., \& Adelman, S. J. 1994, \apj, 429, L81

\bibitem[]{} Hanner, M. S., Tokunaga, A. T., Veeder, G. J., \& Ahearn, M. F. 1984, \aj, 89, 162

\bibitem[]{} Hawarden, T. G., Leggett, S. K., Letawsky, M. B., Ballantyne, D. R., \& Casali, M. M. 2001, \mnras, 325, 563

\bibitem[]{} Koornneef, J., Bohlin, R., Buser, R., Horne, K., \& Turnshek, D. 1986, in Highlights in Astronomy, ed. J.-P. Swings (Dordrecht: D. Reidel), 833

\bibitem[]{} Landolt, A. U. 1992, \aj, 104, 340

\bibitem[]{} Leggett, S. K., Bartholomew, M., Mountain, C. M., \& Selby, M. J. 1986, MNRAS, 223, 443

\bibitem[]{} Leggett, S. K., et al. 2003, \mnras, 345, 144

\bibitem[]{} Lord, S. D. 1992, NASA Tech. Mem. 103957

\bibitem[]{} M\'{e}gessier, C. 1995, \aap, 296, 771


\bibitem[]{} Oke, J. B., \& Gunn, J. E. 1983, \apj, 266, 713

\bibitem[]{} Persson, S. E., Murphy, D. C., Krzeminski, W., Roth, M., \& Rieke, M. J. 1998, \aj, 116, 2475

\bibitem[]{} Peterson, D. M., et al. 2004. in Proc. SPIE, 5491, 65

\bibitem[]{} Price, S. D., Paxson, C., Engelke, C., \& Murdock, T. L. 2004, \aj, 128, 889

\bibitem[]{} Selby, M. J., Mountain, C. M., Blackwell, D. E., Petford, A. D., \& Leggett, S. K. 1983, \mnras, 203, 795

\bibitem[]{} Simons, D. A., \& Tokunaga, A. 2002, \pasp, 114, 169

\bibitem[]{} Stephens, D. C., \& Leggett, S. K. 2004, \pasp, 116, 9

\bibitem[]{} Tokunaga, A. T., Simons, D. A., \& Vacca, W. D. 2002, \pasp, 114, 180

\bibitem[]{} Wamsteker, W. 1981, \aap, 97, 329

\bibitem[]{} Young, A. T., Milone, E. F., \& Stagg, C. R. 1994, \aaps, 105, 259


\end{thebibliography}
\end{document}